\documentclass[prb,twocolumn,floats,graphicx,superscriptaddress]{revtex4}
\usepackage{graphicx}
\newcommand{\beq}{\begin{equation}}
\newcommand{\eeq}{\end{equation}}
\newcommand{\bqn}{\begin{eqnarray}}
\newcommand{\eqn}{\end{eqnarray}}

\begin{document} \title{Silent Phase Qubit Based on $d$-Wave
Josephson Junctions }

\author{M.H.S. Amin}
\affiliation{D--Wave Systems Inc., 320-1985 Broadway, Vancouver,
B.C., V6J 4Y3, Canada}

\author{A.Yu.~Smirnov}
\affiliation{D--Wave Systems Inc., 320-1985 Broadway, Vancouver,
B.C., V6J 4Y3, Canada}

\author{A.M.~Zagoskin}
\affiliation{D--Wave Systems Inc., 320-1985 Broadway, Vancouver,
B.C., V6J 4Y3, Canada} \affiliation{Physics and Astronomy Dept.,
The University of British Columbia, 6224 Agricultural Rd.,
Vancouver, B.C., V6T 1Z1, Canada}

\author{T.~Lindstr\"{o}m}
\affiliation{Microtechnology and Nanoscience, Quantum Device
Physics Laboratory, Chalmers University of Technology and
G\"oteborg University, SE-412 96 G\"oteborg, Sweden}

\author{S.~Charlebois}
\affiliation{Microtechnology and Nanoscience, Quantum Device
Physics Laboratory, Chalmers University of Technology and
G\"oteborg University, SE-412 96 G\"oteborg, Sweden}

\author{T.~Claeson}
\affiliation{Microtechnology and Nanoscience, Quantum Device
Physics Laboratory, Chalmers University of Technology and
G\"oteborg University, SE-412 96 G\"oteborg, Sweden}

\author{A.Ya.~Tzalenchuk}
\affiliation{National Physical Laboratory, Queens Road,
Teddington, Middlesex TW11 0LW, UK}

\begin{abstract}

We report on design and fabrication of a new type of flux qubit
that capitalizes on intrinsic properties of submicron YBCO grain
boundary junctions.  The operating point is protected from the
fluctuations of the external fields, already on the classical
level; the effects of external perturbations are absent, to the
second or third order, depending on the character of the
coupling. We propose an experiment to observe quantum tunneling
and Rabi oscillations in the qubit. Estimate of the decoherence
due to fluctuations of the external flux is presented.

\end{abstract}

\maketitle


Over the last few years a series of experiments
\cite{expt,mooij,vion} provided a conclusive evidence of quantum
superposition in meso- and macroscopic superconductors. The task
at hand is scaling up of the system, with two goals in mind: to
probe how far quantum superposition can be pushed into
macroscopic world, and to develop an element base for quantum
computing, which only becomes viable on the scale 10--100 qubits.

It was suggested, that use of high-T$_c$ cuprates in qubits would
dispose of the need to apply fine-tuned external fields to keep
it in the operating point, due to the time-reversal symmetry
breaking in systems with DD junctions\cite{Blatter,zagoskin}. On
the level of a few qubits this is not a major advantage, compared
to the relative difficulty of fabrication and threat of extra
decoherence from nodal quasiparticles and zero-energy states (ZES)
in cuprates. Therefore the research was concentrated on
conventional superconductors, where all of the aforementioned
successes were achieved.

\begin{figure}[t]
\includegraphics[width=8cm]{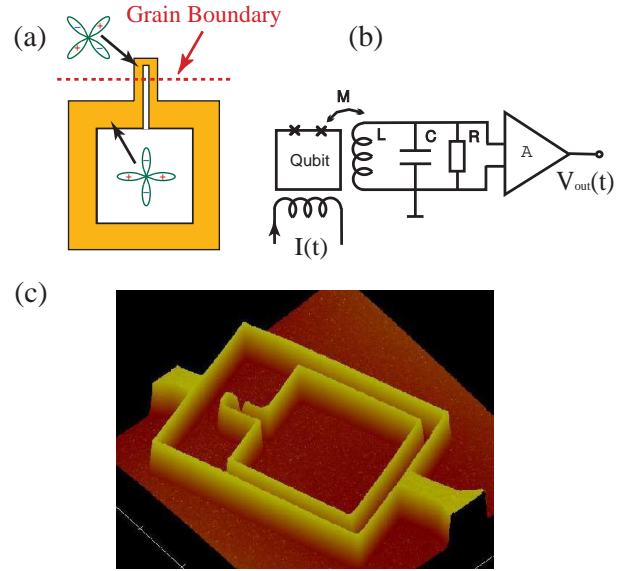}
\caption{{\protect\small (a) ``Silent" qubit. The dashed line
shows the location of the grain boundary junction. (b) Suggested
scheme for observation of Rabi oscillation and measurement of
decoherence. (c) AFM picture of a fabricated qubit with a
dc-SQUID fabricated around it for readout of the states.}}
\label{fig1}
\end{figure}

Nevertheless, the logical progression of research from single
qubit to qubit-qubit coupling and further to qubit registers is
bringing us to the point where intrinsic bistability of
high-T$_c$ qubits will become a major advantage. In the last few
years, we saw the development of reliable fabrication of bistable
submicron high-T$_c$ structures\cite{Ilichev01,Ivanov,Lindstrom}.
The danger of nodal quasiparticles and ZES seems to be
overestimated\cite{AminSmirnov,Fominov}. Finally, putting aside
any technological motivation, it is tempting to probe high-T$_c$
compounds for the same kind of macroscopic quantum coherence as
was observed in low-T$_c$ superconductors, just to see how and
whether these phenomena depend on, e.g., symmetry of the ground
state of the system.

The ``silent qubit" (for the etimology, see below) can be viewed
as a SQUID formed out of high-T$_c$ film, with two DD-grain
boundary junctions forming a mesoscopic island
(Fig.~\ref{fig1}a,c). The crystal lattice and $d$-wave order
parameter orientations on the two sides of the grain boundary are
chosen in such a way that each junction has doubly degenerate
ground state \cite{Ilichev01,amin}.

To simplify the analysis, we keep only two harmonics in the
current-phase relations of the DD junctions: $ I_{i}=I_{ci}\sin
\phi _{i}-I_{ci}^{\prime }\sin 2\phi _{i}$, where $I_{ci}$ is the
Josephson current, and $\phi _{i}$ is the phase difference across
the $i^{{\rm th}}$ junction ($i=1,2$). This approximation
successfully reproduces the observed behaviour of similar devices
in classical regime\cite{Ilichev01,Lindstrom}. The flux
quantization condition binds the phases $\phi _{1}$ and $\phi
_{2}$   to the total flux through the loop $\Phi$ via $\phi
\equiv \phi_{1}+\phi_{2}=2\pi \Phi /\Phi_{0}$, where $\Phi
_{0}=h/2e$ is the flux quantum. Introducing the superconducting
phase of the island $\theta \equiv (\phi _{1}-\phi _{2})/2$, the
free energy of the qubit, threaded by an external magnetic flux
$\Phi_x$,  in units of the Josephson energy $E_J$ ($\equiv E_1$,
where $E_{i}=I_{ci}\Phi_0/2\pi$), is given by
\begin{equation}
{\cal U}(\theta ,\phi )
= {\frac{(\phi -\phi _{x})^{2}}{2{\beta_L}}} -{\cal E}_{\phi
}\left[ \cos \theta -\frac{\tilde{\alpha}_\phi}{4} \cos (2\theta
)\right] +\tilde{\cal U}(\theta ,\phi ). \label{Ut}
\end{equation}
Here   $\alpha _{i} = 2I_{ci}^{\prime}/I_{ci},$   ${\cal E}_{\phi
}=(1+\eta )\cos (\phi/2), \tilde{\alpha}_\phi= \left[(\alpha
_{1}+\eta \alpha _{2})\cos \phi\right]/\left[(1+\eta) \cos \left(
\phi /2\right)\right],$ $\eta =E_{2}/E_{1}{,}$ and $\phi _{x}=2\pi
\Phi _{x}/\Phi _{0}.$ The dimensionless self-inductance of the
loop ${\beta_L} \equiv 2\pi LI_{c1}/\Phi_0$ is considered
negligible (${\beta_L}\to 0$); then $\phi \to \phi _{x}$ and the
first term in (\ref{Ut}) can be dropped.

The last term in (\ref{Ut}):
$$
\tilde{\cal U}(\theta ,\phi )=-\left[ \eta -1+(\alpha _{1}-\eta
\alpha _{2})\cos \frac{\phi }{2}\cos \theta \right] \sin
{\frac{\phi }{2}}\sin \theta ,
$$
is zero when the two junctions are identical (i.e. $\eta =1$ and $%
\alpha _{1}=\alpha _{2}$). In this case the potential energy
minima are at $\theta = \pm \arccos (1/\tilde{\alpha}_\phi) $ when
$\tilde{\alpha}_\phi>1$ and zero otherwise. The $\pm$ signs
correspond to the states on the right and left sides of the
potential well (i.e. $|\pm\rangle$), respectively. For
$\tilde{\alpha}_{\phi}>1$ the potential has two minima, which are
degenerate at {\em any} external flux $\phi _{x}.$  The current
induced in the loop by the external flux does {\em not} depend on
the state of the qubit, which justifies the moniker ``silent". The
potential profile (\ref{Ut}) is similar to the one of persistent
current qubit \cite{mooij}, but here we have only one independent
phase (as ${\beta_L} \to 0$), and the problem becomes
one-dimensional. The barrier between the potential minima is
flux-dependent: $ W={\cos (\phi_x /2)}(
\tilde{\alpha}_\phi+\tilde{\alpha}_\phi^{-1}-2).$

In the general case ($\alpha_1\neq\alpha_2,$ $\eta\neq 1$) the
two minima are only degenerate when $\phi _{x}=0$. Now a {\em
state-dependent} persistent current flows in the loop even in
zero external field; in units of $I_{c1}$, ${\cal I}_{0}^{\pm
}=\pm [\eta \sqrt{\tilde{\alpha}_0^{2}-1}/(1+\eta
)\tilde{\alpha}_0^{2}](\alpha _{2}-\alpha _{1})$,
%
where $\tilde{\alpha}_0 \equiv \tilde{\alpha}_{\phi =0}$.

An intermediate regime ($\alpha_1=\alpha_2$, $\eta\neq 1$) is most
interesting. It takes place when junctions have similar
current-phase dependences, but different critical currents (i.e.
widths), and should be expected if the junctions are close enough
to each other (Fig.~\ref{fig1}). At $\phi_x=0$, the equilibrium
value of $\theta$ is the same for both junctions, and there is no
spontaneous current: ${\cal I}_{0}^{\pm }=0$. At finite $\phi_x$,
the induced currents differ  for the two states of the qubit, but
the difference is of higher order in $\phi_x$, keeping the qubit
silent.

\begin{figure}[t]
\includegraphics[width=8.7cm]{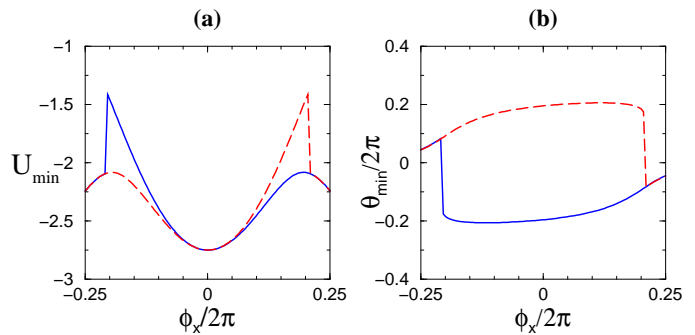}
\caption{{\protect\small The minimum of the free energy (a) and
the corresponding values of $\theta$ at the minimum points (b) as
a function of $\phi_x$, for $\eta=2$ and $\alpha_1=\alpha_2=3$.
The solid and dashed lines correspond to the $\left|-\right>$ and
$\left|+\right>$ states respectively.}} \label{fig2}
\end{figure}

Expanding the free energy   to the third order in $\phi _{x}$, we
find for the minima:
\begin{equation}
{\cal U}_{{\rm min}}^{\pm }=A_{0}+A_{2}\phi _{x}^{2}\pm A_{3}\phi
_{x}^{3} \label{Umin}
\end{equation}
where $A_i$ are explicit, but cumbersome, functions of
$\tilde{\alpha}_0$ and $\eta$. As expected, there is no first
order dependence on $\phi_{x}$ ($A_{1}=0 $). The second order
term in (\ref{Umin}) does not depend on the state of the qubit.
The first state-dependent term in the minimum energy of the
system is $O(\phi _{x}^3)$ (Fig.~\ref{fig2}). Therefore small
fluctuations of $\phi _{x}$ do not affect the degeneracy of the
states. The difference between the energy minima grows as the
external flux is increased until the point at which the potential
barrier vanishes altogether, and the minimum with higher energy
disappears (the jumps in Fig.~\ref{fig2}).

\begin{figure}[t]
\includegraphics[width=55mm]{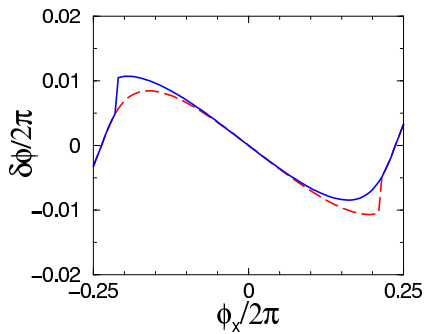}
\caption{\protect\small Self-consistent calculation of the
self-generated flux for the system of Fig.~\ref{fig2} with
${\beta_L}=0.01$. } \label{fig3}
\end{figure}

The current in the loop is found from $ {\cal I}_{\phi }^{\pm } =
\partial_{\phi_x} {\cal U}_{\rm min}^\pm = 2A_{2}\phi _{x}\pm
3A_{3}\phi _{x}^{2}$. This current generates a state-dependent
magnetic flux $\delta \phi \equiv \phi -\phi _{x}= {\beta_L}{\cal
I}_{\phi }^{\pm }$. Note that the state-dependent contribution to
the induced flux is $O(\phi_x^2).$ For finite self-inductances
${\beta_L}$, $\delta \phi$ has to be calculated
self-consistently. Figure~\ref{fig3} shows the result of such a
calculation. With the parameters chosen, an external flux close
to $0.2\Phi _{0}$ generates an additional flux ($\sim 0.005\Phi
_{0}$) through the loop, which is of the same order as the
estimate based on the above expansion.

Tunneling between the potential minima occurs due to the
uncertainty relation between the charge $Q$ of the island and its
superconducting phase $\theta$.  The tunneling matrix element is
approximately given by ($\hbar=k_{\rm B}=1$) $ \Delta \approx
\omega_p(\phi_x) e^{-\sqrt{\zeta W(\phi_x) {E_J/ E_c} }}$, where
$\zeta$ is a constant of the order of 1. The coefficient
$\omega_p(\phi_x) \equiv \sqrt{\omega_p^+\omega_p^-}$ is
determined by the frequencies $\omega_p^\pm$ of small oscillations
in the right and left potential minima, respectively. This
dependence follows from the expression of $\Delta$ as the matrix
element between the lowest energy states in the two wells. In the
case of $\tilde{\alpha}_0 \sim 1$, it is only valid
qualitatively\cite{TZA}, but enough for the present analysis. Due
to the symmetry of the potential profile when
$\alpha_1=\alpha_2$, the linear dependence on $\phi_x$ cancel and
we are left with $\omega_p(\phi_x) = \sqrt{E_J E_C (\alpha_0 -
\alpha_0^{-1})} (1-\kappa\phi_x^2), $ where $E_C = e^2 /2C$ is
the charging energy, $C$ is the effective capacitance of the
junctions, and $\kappa $ is a dimensionless coefficient of $O(1)$.
Fluctuations of $\phi_x$ influence $W$ and therefore $\Delta$.
Expanding the Josephson potential near the origin, we obtain the
tunnelling barrier $W={\cal U}_{\max }-{\cal U}_{\min }^{\pm
}=(B_{0}-A_{0})+\left( B_{2}-A_{2}\right) \phi _{x}^{2} +
O(\phi_x^3) $, where $B_{0} =-\left( \eta +1\right) \left(
1-\tilde{\alpha}_0/4\right),$ $B_{2} = -\left[(\tilde{\alpha}_0
-1)/4\right]\left[\eta/(\eta +1)\right]$. Again there is no
dependence on $\phi_x$ in the lowest order.

Truncating the Hilbert space of the qubit to the two lowest energy
states, one can write the effective Hamiltonian of the system as $
H=\frac{1}{2}\Delta (\phi _{x})\sigma _{x}+\frac{1}{2}\epsilon
(\phi _{x})\sigma _{z}$, where $\epsilon (\phi _{x})\approx
E_{J}A_{3}\phi _{x}^{3}$. All single qubit operations can be
realized by applying controlled flux $\phi _{x}.$ Note that the
qubit only leaves the operating point when a finite external flux
is applied. Unlike the earlier qubit designs\cite{mooij,vion} this
point is protected from external flux fluctuations, already on
the classical level [cf. Eq.~(\ref{Umin})].

Readout of the quantum state can in principle be achieved using a
dc-SQUID to measure the magnetic flux generated by the qubit when
it is biased by an external magnetic field. In this work the
readout SQUID was fabricated onchip using a high-$T_c$ grain
boundary SQUID (cf. Fig.~\ref{fig1}c); alternatively it can be
made of aluminum. Although coherent oscillations in an
all-aluminum persistent current qubit have previously been
observed using a similar readout design\cite{mooij}, the measured
decoherence time appeared to be rather short. Subsequent results
of Il'ichev et al.\cite{Rabi} have demonstrated a longer
decoherence time in a measurement scheme which couples the qubit
to a high quality tank circuit. Here, we calculate the decoherence
time of our silent qubit in the same setup\cite{Grajcar,Rabi}
(see Fig.~\ref{fig1}b). The circuit is assumed to have a
resonance frequency $\omega_T$ and the damping rate $\gamma_T$.
The mutual inductance is $M = k\sqrt{L_T L}$, $k \leq 1$ being
the coupling coefficient, and $L$ and $L_T$ are the inductance of
the qubit and tank respectively.

Without limiting the generality of our approach, we can ascribe
all the dephasing and dissipation in the qubit to its interaction
with the fluctuating flux $\phi_x(t)$, created by the tank current
$I_T(t)$. These fluctuations are characterized by correlator
$K(t,t')= \langle I_T(t), I_T(t') \rangle$, spectral density
$
K(\omega ) = \omega (\gamma_T \omega_T^2 / L_T) [\coth(\omega/2T) + 1] / [(\omega^2 - \omega_T^2)^2 +
\gamma_T^2 \omega^2 ],$
and dispersion $\langle I_T^2\rangle = K(t,t)$ $= ( \omega_T /
2L_T) \coth (\omega_T /2T).$ At $\phi_x=0$, the qubit Hamiltonian
becomes $ H = (\Delta /2) \sigma_x  - \lambda_3 I_T^3
\sigma_z - \lambda_2 I_T^2 \sigma_x$, where $\lambda_3$ and
$\lambda_2$ are coupling coefficients depending on the qubit
parameters. In the Bloch-Redfield approximation\cite{smirnov}, we
calculate the energy relaxation rate $\Gamma = 30 \lambda_3^2
\langle I_T^2\rangle^2 (\gamma_T / \Delta^3) (\omega_T^2 /
L_T) \coth(\Delta /2T)$, together with a
dephasing time of the qubit $\gamma^{-1},$ where $ \gamma =
\Gamma/2 + \gamma_0 $,  $ \gamma_0 = (16 \pi / 3) \lambda_2^2
(\gamma_T^2 T^3 / L_T^2 \omega_T^4),$ if $T \ll \omega_T,$ and
$\gamma_0 = \lambda_2^2 Q_T (\omega_T / L_T^2) [
\coth^2(\omega_T / 2T) - 1 ] $ at
temperatures $T\geq \omega_T.$

Using the experimental data of Ref.~\onlinecite{TZA} ($I_c=0.5$
$\mu$A, $I_c'=0.6$ $\mu$A, $C\approx 10$ fF), we find $E_C \approx
2$ GHz, $\omega_p/2\pi \approx 40$ GHz, and $\Delta/2\pi \approx
1.6$ GHz. For ${\beta_L}\sim 0.01$, which is the value used in
Fig.~\ref{fig3}, the inductance of the loop is of the order of
$10pH$. To estimate the contributions of the cubic and quadratic
terms to qubit dephasing and dissipation, we chose the following
parameters: $\eta = 2,\ \alpha_1 = \alpha_2 = 2.4$, and $E_J
=1.66\times 10^{-22} J$.  If the tank frequency $\omega_T/2\pi =
10$ MHz, its quality factor $Q_T = 2000,$ and the coupling
coefficient $ k \sim 1/33$, then contribution of {\em quadratic}
flux fluctuations to dephasing rate is small, so that the
dephasing time due to  qubit coupling to the tank is
$\gamma_0^{-1} \simeq 20$ ms at temperatures of order $10$ mK,
while the contribution of the {\em cubic} fluctuations to the
dephasing and relaxation rates is totally negligible. It means
that at the operating point the silent qubit is practically
decoupled from the fluctuations caused by the controlling
circuits. The dominant source of decoherence is from the nodal
quasiparticles at the junction, considered in
Ref.~\onlinecite{AminSmirnov}, which may reduce the decoherence
time to about 1--100 ns.

In conclusion, a new type of flux qubit, based on specific
properties of submicron YBCO grain boundary junctions, is proposed
and fabricated. The symmetry of the device provides an operating
point, which is stable and protected against the external field
fluctuations.

We are grateful to A. Golubov, E. Il'ichev, A. Maassen van den
Brink, Y. Nakamura, and A. Shnirman for stimulating discussions.



%
%



\end{document}